\documentclass[aps,prl,twocolumn,groupedaddress]{revtex4-1}
\usepackage{amsmath,amssymb,color}
\usepackage{epsfig}
\begin{document}
\def\E{\mathbb{E}}
\def\P{\mathbb{P}}
\def\R{\mathbb{R}}
\def\Z{\mathbb{Z}}
\def\O{{\cal O}}
\def\scr{\scriptstyle}
\def\text{\textstyle}
\def\floor{\rm floor}
\def\sw{\rm sw}
\def\al{\alpha}
\def\be{\beta}
\def\ga{\gamma}
\def\del{\delta}
\def\la{\lambda}
\def\sig{\sigma}
\def\om{\omega}
\def\Om{\Omega}
\def\phi{\varphi}
\def\na{\nabla}
\def\A{{\cal A}}
\def\atanh{\rm atanh}
\def\tiz1{\tilde z_1}
\def\p{\partial}
\def\sub{\subset}
\def\12{\frac{1}{2}}
\def\oraw{\overrightarrow}
\def\nea{\nearrow}
\def\sea{\searrow}
\def\beq{\begin{equation}}
\def\eeq{\end{equation}}
\def\beqn{\begin{equation*}}
\def\eeqn{\end{equation*}}
\def\beqa{\begin{eqnarray}}
\def\eeqa{\end{eqnarray}}
\def\beqan{\begin{eqnarray*}}
\def\eeqan{\end{eqnarray*}}
\title{Contact angles of a drop pinned on an incline}
\author{Jo\"el De Coninck}
\affiliation{Laboratoire de Physique des Surfaces et Interfaces\\
Universit\'{e} de Mons, 20 Place du Parc, 7000 Mons, Belgium}
\author{Fran\c cois Dunlop} \author{Thierry Huillet}
\affiliation{Laboratoire de Physique Th\'{e}orique et Mod\'{e}lisation,
CNRS-UMR 8089\\ Universit\'{e} de Cergy-Pontoise, 
95302 Cergy-Pontoise, France}
\begin{abstract}
For a drop on an incline with small tilt angle $\al$, when the contact line
is a circle of radius $r$, we derive the relation
$mg\sin\al=\ga r{\pi\over2}(\cos\theta^{\rm min}-\cos\theta^{\rm max})$ at first
order in $\al$, where $\theta^{\rm min}$ and $\theta^{\rm max}$ are the contact 
angles at the back and at the front, $m$ is the mass of the drop and $\ga$ the 
surface tension of the liquid. We also derive the same relation at first order
in the Bond number $B=\rho g R^2/\ga$, where $R$ is the radius of the spherical 
cap at zero gravity. The drop profile is computed exactly in the same 
approximation. These results are compared with {\sl Surface evolver} results,
showing a surprisingly large range of applicability of first order 
approximations. 
\end{abstract}
\maketitle
Pinning of a drop on an incline is a subject with a long history, see
\cite{MO}\cite{Fr} and \cite{BT} and references therein.
Hysteresis of contact angle and the roll-off angle have attracted renewed
attention in recent years, motivated by the search for super-hydrophobic
materials. The roll-off angle 
is often used to characterize the quality of a surface : if it is small, say 
below 5$^\circ$, the surface is considered as perfect such as a piece of glass or
silica wafer. Small roll-off angles are also observed with super-hydrophobic
materials such as the lotus leaf. If the roll-off angle is large, say above 
10$^\circ$, then the surface must be heterogeneous, physically in terms of 
topographical defects or chemically in terms of various species covering the 
surface, or for most of the industrial cases, both.
\begin{figure}
\begin{center}
\resizebox{7cm}{!}{\includegraphics{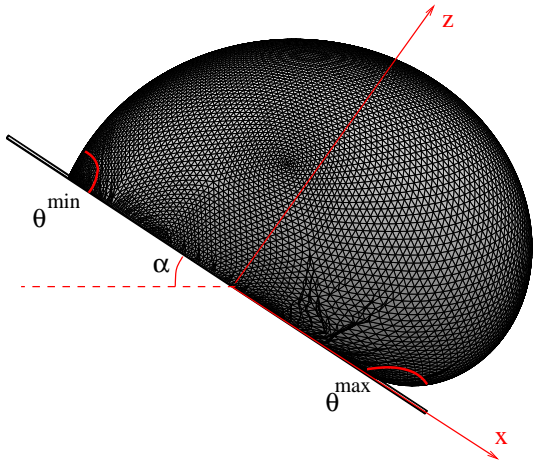}}
\caption{Water drop on hydrophobic incline at angle $\al=30^\circ$ . Volume 
$V\simeq42\mu$L. Pinned base radius $r_0\simeq 2.0\,$mm. 
Simulated with {\sl Surface Evolver}.}
\label{2pi3B07}
\end{center}
\end{figure}
Advancing and receding contact angles may be defined as follows (see \cite{dGBQ}
for background and references): consider
a small piece of contact line where the three phases meet. The sum of forces
parallel to the solid surface, per unit length of contact line, is perpendicular
to the line and defines the local spreading coefficient
$\ga_{SV}-\ga_{SL}-\ga\cos\theta=\ga\,(\cos\theta_Y-\cos\theta)$
where $\theta$ is the local contact angle and $\theta_Y$ is the Young angle
implied by the equation. The local contact angle $\theta$ is a macroscopic 
quantity, with smooth variation on the macroscopic scale, because the fluid 
surface is smooth. The Young angle $\theta_Y$, before averaging, follows the 
heterogeneity of the solid surface energies $\ga_{SV}-\ga_{SL}$ and may vary 
in a range $\theta_1\le\theta_Y\le\theta_2$. If the local contact angle falls
in this range then the piece  of contact line will undergo positive and negative
spreading coefficients and thus will be pinned. Otherwise it will move to one
side or the other, defining advancing and receding contact angles. 

This is a simplified picture, notably because it deals with metastability 
through equlibrium macroscopic notions only, which 
will be wrong at the nano-scale. One should also distinguish Wenzel states 
wetting nano-pores from Cassie-Baxter states with air pockets, etc. The 
advancing and receding angles $\theta^A$  and $\theta^R$ are defined 
experimentally. But the basic mechanism should be valid, and should imply the 
following scenario: a drop is gently deposited on a horizontal substrate; 
the macroscopic contact line is a circle.
Suppose the contact angle is  $\theta^R<\theta_0<\theta^A$. Now tilt the
substrate by a small angle $\al$. The contact angle along the contact line
becomes a function $\theta(\phi)$ oscillating around $\theta_0$ and therefore 
satisfying $\theta^R<\theta(\phi)<\theta^A$ for all $\phi$. The contact line
is pinned everywhere and remains circular. Upon increasing $\al$, depending upon
$\theta_0$, the maximum of $\theta(\phi)$ will reach $\theta^A$ or the minimum
will reach $\theta^R$ and a corresponding piece of the contact line will move by
a finite amount. The remaining piece holds the drop. Upon increasing $\al$ 
further, eventually the remaining piece will be unable to hold the drop, or it 
will reach its limit $\theta^R$ or $\theta^A$ and the drop will roll off. Such a
scenario with three different transitions was already proposed in \cite{BT}. 
If $\theta_0=\theta^A$ or $\theta^R$, of course the first stage is skipped,
and the circle is deformed as soon as $\al>0$. The importance of the deposition 
history was already stressed in \cite{MSKK}\cite{CHST}\cite{WSRV}.

In this letter we consider the first stage, where the contact line is
pinned as a circle of radius $r_0$. We denote  $\theta^{\rm max}_\al$ and 
$\theta^{\rm min}_\al$ the contact angles at the front and at the back of the drop 
when the tilt angle is $\al$ (see Fig. \ref{2pi3B07}). 
We show, for any $B$, for small $\al$,
\beq\label{mga}
mg\sin\al= \ga r_0{\pi\over2}\Bigl(\cos\theta^{\rm min}_\al-\cos\theta^{\rm max}_\al
\Bigr)+\O(\al^3)
\eeq
and for any $\al$, for small $B$,
\beq\label{mgaB}
mg\sin\al= \ga r_0{\pi\over2}\Bigl(\cos\theta^{\rm min}_\al-\cos\theta^{\rm max}_\al
\Bigr)+\O(B^2)
\eeq
Our derivations are analytic, but a factor $\pi/2$ or very near $\pi/2$
was found previously from numerical solutions using the finite elements method
\cite{BOS} or from experiments \cite{EJ04I}\cite{EJ06}, see Fig. 4 in 
\cite{EJ06}. We use {\sl Surface evolver} \cite{Br} to compare first order
approximations and numerically almost exact results. 

We start from a sessile drop on a plane horizontal 
substrate, with three-phase contact-line a circle of radius $r_0$.
We use cylindrical coordinates $(z,r,\phi)$ 
with origin at the center of the contact-line circle. The hydrostatic pressure 
just below the drop surface is $p=p_0-\rho g z$
where $p_0$ is the pressure at the origin and $z=z(r)$ is the drop profile,
obeying the Laplace-Young equation
\beqn
p-p_{\rm atm}=-2\ga H
=-\ga\Bigl({z''\over(1+z'^2)^{3/2}}+{z'\over r\,(1+z'^2)^{1/2}}\Bigr)
\eeqn 
where $\ga$ is the liquid-air surface tension and $H$ is the
mean curvature. The boundary conditions are $z'(0)=0\,,\ z(r_0)=0$.
Eliminating the pressure gives
\beq\label{ode}
p_0-p_{\rm atm}=\rho g z-2\ga H
\eeq 
The parameters $r_0$ and $p_0-p_{\rm atm}$ may be changed in terms of drop volume
and macroscopic contact angle $\theta_0$. This angle depends upon the way the
sessile drop was deposited on the substrate, and can be any angle between the
receding angle and the advancing angle.

Let us now tilt the substrate by an angle $\alpha$, and assume that the contact 
line does not move, as discussed above. 
We keep cylindrical cooordinates with $z$-axis perpendicular to the substrate,
so that the hydrostatic pressure is now
\beqn
p=p_0-\rho g z\cos\al+\rho g x\sin\al
\eeqn
where the $x$-axis is chosen in the direction of the downward slope. Then 
(\ref{ode}) becomes
\beq\label{odex}
p_0-p_{\rm atm}=\rho g z\cos\al-\rho g x\sin\al-2\ga H
\eeq 
where now $z=z(r,\phi)$, with partial derivatives denoted 
$z_r,z_\phi,z_{rr},z_{r\phi},z_{\phi\phi}$, and
\beqa\label{kphi}
2H=\Bigl( r^2\bigl(z_r^2+1\bigr)+z_{\phi\phi}^2\Bigr)^{-3/2}\Bigl[
rz_{rr}\bigl(z_\phi^2+r^2\bigr)+\hskip1cm\cr
z_rr^2\bigl(z_r^2+1\bigr)+2z_rz_\phi\bigl(z_\phi-rz_{r\phi}\bigr)
     +rz_{\phi\phi}\bigl(z_r^2+1\bigr)\Bigr]\hskip0.5cm
\eeqa

At small tilt or small Bond number the solution to (\ref{odex}) will generally
admit a Taylor expansion in a small parameter, and one may attempt to solve
(\ref{odex}) order by order. We consider the first order, which corresponds to
linearizing (\ref{odex}). We assume that order zero has cylindrical symmetry,
so that $z(r,\phi)=z_0(r)+\al z_1(r,\phi)+{\rm higher\ orders}$, or a similar
formula with the Bond number instead of $\al$, and the appropriate $z_0$ in each
case. Inserted into (\ref{kphi}) this
yields $H=H_0+\al H_1+{\rm higher\ orders}$ or a similar formula with 
the Bond number instead of $\al$, with, in any case,
\beqa\label{H1}
2H_1=\bigl(1+z_0'^2\bigr)^{-{3\over2}}z_{1rr}+(1+z_0'^2)^{-\12}\,{z_{1\phi\phi}\over r^2}
+\cr
(1+z_0'^2)^{-{3\over2}}{z_{1r}\over r}
-3z_0''z_0'(1+z_0'^2)^{-{5\over2}}z_{1r}
\eeqa
Volume conservation and boundary conditions apply to all orders. In particular,
\beqn
0=\int_0^{2\pi}d\phi\int_0^{r_0}dr\,r\,z_1(r,\phi)\,,\qquad 
z_1(r_0,\phi)=0\quad\forall\ \phi
\eeqn

\subsection{Small tilt}\label{tilt}
Here we take for $z_0$ the solution of (\ref{ode}). The pressure $p_0$
at the center is even in $\al$, so that $p_0=p_{00}+\O(\al^2)$. Order zero in
(\ref{odex}) is (\ref{ode}) now written as
\beqn
p_{00}-p_{\rm atm}=\rho g z_0-2\ga H_0
\eeqn 
and the contact angle at order zero is given by $\tan\theta_0=-z_0'(r_0)$.
Order one, the coefficient of $\al$ in the Taylor expansion of (\ref{odex})
with $z(r,\phi)=z_0(r)+\al z_1(r,\phi)+{\rm higher\ orders}$, is
\beq\label{odex1}
0=\rho gz_1-\rho g r\cos\phi-2\ga H_1
\eeq 
where the polar angle $\phi$ is measured from the downward slope direction.
An ansatz for a solution is
\beqn
z_1(r,\phi)=\tilde z_1(r)\cos\phi\,,\qquad\tiz1(0)=0,\quad \tiz1(r_0)=0
\eeqn
Then using (\ref{H1}) it appears that $\cos\phi$ cancels out from (\ref{odex1}),
and $\tilde z_1(r)$ is the solution of the ordinary differential equation
\beqa\label{ode1}
{\rho gr\over\ga}={\rho g\tiz1\over\ga}-\bigl(1+z_0'^2\bigr)^{-{3\over2}}\tiz1''
-(1+z_0'^2)^{-\12}\,{\tiz1\over r^2}\cr
+(1+z_0'^2)^{-{3\over2}}{\tiz1'\over r}-3z_0''z_0'(1+z_0'^2)^{-{5\over2}}\tiz1'
\eeqa
The contact angle $\theta_\al(\phi)$ obeys
\beqn
\tan\theta_\al(\phi)=-{\p z\over\p r}(r_0,\phi)=\tan\theta_0
-\al\tiz1'(r_0)\cos\phi+\O(\al^2)
\eeqn
so that
\beq\label{thetaphi}
\cos\theta_\al(\phi)=\cos\theta_0
+\al\tiz1'(r_0)\sin\theta_0\cos^2\theta_0\cos\phi+\O(\al^2)
\eeq
and 
\beq\label{thetaphi2}
{\cos\theta_\al(\phi)-\cos\theta_\al^{\rm max}\over\cos\theta_\al^{\rm min}
-\cos\theta_\al^{\rm max}}={1-\cos\phi\over2}+\O(\al)
\eeq
to be compared to ElSherbini and Jacobi's formula \cite{EJ04I}
\beq\label{EJ}
{\theta_\al(\phi)-\theta_\al^{\rm min}\over\theta_\al^{\rm max}-\theta_\al^{\rm min}}
=2{\phi^3\over\pi^3}-3{\phi^2\over\pi^2}+1
\eeq
A comparison is achieved by plotting the right-hand-side of (\ref{EJ}) together 
with the function of $\phi$ obtained from the left-hand-side of (\ref{EJ}) with
$\theta_\al(\phi)$ extracted from (\ref{thetaphi2}) without $\O(\al)$.
The two plots are hardly distinguishable when 
$\theta_\al^{\rm max}-\theta_\al^{\rm min}$ is small, as considered here. 

The total capillary force upon the drop, projected onto the substrate
and onto the direction $\phi=\pi$, upwards along the slope, is:
\beqa
F_\ga&=&-\ga r_0\int_{-\pi}^\pi d\phi\cos\phi\cos\theta_\al(\phi)\cr
&=&\ga r_0{\pi\over2}\Bigl(\cos\theta_\al^{\rm min}-\cos\theta_\al^{\rm max}\Bigr)
+\O(\al^3)
\eeqa
\begin{figure}
\begin{center}
\vskip-2.0cm
\resizebox{8.5cm}{!}{\includegraphics{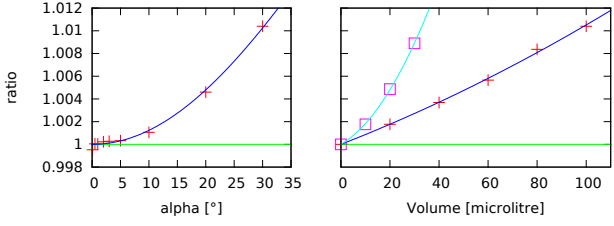}}
\caption{The ratio (\ref{ratio}) 
function of $\al$ (left: $100\mu$L droplet, with fit 
$1+c\sin^2\al$), and function of drop volume (right: +($\al=30^\circ$) and 
$\square$($\al=60^\circ$), each with fit $1+aV+bV^2$).}\label{gg}
\end{center}
\end{figure}
The error is $\O(\al^3)$ because the part even in $\al$ cancels out when 
integrating over $\phi$.
Equilibrium with gravity implies $F_\ga=mg\sin\al$, giving (\ref{mga}), implying
\beq\label{ratio}
{\ga r_0\pi(\cos\theta_\al^{\rm min}-\cos\theta_\al^{\rm max})\over 2mg\sin\al}
=1+\O(\al^2)
\quad{\rm as}\ \al\to0 
\eeq
Formula (\ref{thetaphi}) can then be written as
\beqn
\cos\theta_\al(\phi)=\cos\theta_0-{mg\sin\al\over\ga r_0\pi}\cos\phi+\O(\al^2)
\eeqn

We have used {\sl Surface evolver} to compute the ratio 
(\ref{ratio}) numerically for $\al$ varying between 0.1$^\circ$ and 30$^\circ$
for a $100\mu$L droplet with base radius $6\,$mm, correponding to
$\theta_0\simeq 32^\circ$, see Fig. \ref{gg} (left).
Maximum and minimum contact angles, in the plane of symmetry of the drop,
were measured by a quadratic fit with three points nearest to the contact line.
The error on the ratio is inversely proportional to the number of vertices
times $\sin\al$. For $\al$ greater than $5^\circ$ error bars are too small to be
shown. For $\al$ smaller than $5^\circ$ the limiting value 1 or a value
derived from the quadratic fit  are better. The value 0.9995 for $\al=0.1^\circ$ 
compared to 1.00003 for $\al=1^\circ$ illustrates the divergence of the error as 
$\al\sea0$.

\subsection{Small Bond number} 
The Bond number is a dimensionless ratio between gravitation and capillarity,
such as $mg/(\ga r_0)$, but more often in the form $\rho gL^2/\ga$, where
popular choices for the length $L$ are $r_0$ or $V^{1/3}$ or $R$, related by
the spherical cap formula,
\beqn
V=\pi R^3\Bigl({2\over3}-\cos\theta_0+{1\over3}\cos^3\theta_0\Bigr)\,,\qquad
r_0=R\sin\theta_0
\eeqn
where $\theta_0$ is the contact angle. All Bond numbers generally give 
the same order of magnitude, but must be specified for quantitative comparisons.
Here we choose $B=\rho gR^2/\ga$ for algebraic simplicity. 
The drop profile at $g=0$ is independent of 
the tilt,
\beq\label{z00}
z_{00}=\sqrt{R^{2}-r^{2}}-\sqrt{R^{2}-r_{0}^{2}}
\eeq
and the corresponding curvature, and the pressure at the origin, are
\beq
H_0=-{1\over R}\,,\qquad p_{00}=p_{\rm atm}+{2\ga\over R}
\eeq
We now assume a Taylor expansion $z(r,\phi)=z_{00}(r)+Bz_1(r,\phi)+\O(B^2)$
which inserted into (\ref{kphi}) yields 
\beq
H=-{1\over R}+BH_1+\O(B^2)
\eeq
with $H_1$ given by (\ref{H1}) with $z_{00}$ instead of $z_0$,
which using (\ref{z00}) simplifies to
\beq\label{g1phi}
2H_{1}=(1-t)^{3/2}z_{1rr}+(1-t)^{1/2}{z_{1\phi\phi}\over r^2}
+(1-t)^{1/2}(1-4t){z_{1r}\over r}
\eeq
where $t=r^2/R^2$.
We then define a dimensionless first order pressure correction $p_1$ by
\beq
p_0-p_{\rm atm}={2\ga\over R}+B{\ga\over R}\,p_1\cos\al+{\rm higher\ orders}
\eeq
Order one in equation (\ref{odex}) takes the form
\beq\label{ode0}
p_1\cos\al={z_{00}\over R}\,\cos\al-{r\over R}\,\sin\al\,\cos\phi-2RH_1
\eeq 
Equation (\ref{odex}) is invariant under $\al\to-\al$, $\phi\to\pi-\phi$, one
can separate odd and even parts of $z-z_{00}$. Accordingly, at first order, we 
try the ansatz
\beq\label{ansatz2}
z_1(r,\phi)=z_{01}(r)\cos\al+z_{11}(r)\sin\al\,\cos\phi
\eeq
Since (\ref{ode0}) is linear and the two terms in (\ref{ansatz2}) are linearly
independent, it yields two independent differential equations where $\cos\al$ 
and $\sin\al\,\cos\phi$ factor out,
\beqa
p_1&=&{z_{00}\over R}-2RH_{01}\,,\ 0=2\pi\int_0^{r_0}dr\,rz_{01},
\ z_{01}(r_0)=0\label{ode000}\hskip0.7cm\\
0&=&-{r\over R}-2RH_{11}\,,\quad z_{11}(0)=0,\quad z_{11}(r_0)=0\label{ode01}
\eeqa
where $2H_{01}$ is (\ref{g1phi}) for $z_{01}$
instead of $z_1$, without the $z_{1\phi\phi}$ term, and
\beqn
2\,H_{11}=(1-t)^{3/2}z_{11}''-(1-t)^{1/2}{z_{11}\over r^2}
+(1-t)^{1/2}(1-4t){z_{11}'\over r}
\eeqn
Like the small tilt case, equations (\ref{ansatz2})(\ref{ode000})(\ref{ode01})
imply (\ref{mgaB}), 
\beq\label{ratioB}
{\ga r_0\pi(\cos\theta_\al^{\rm min}-\cos\theta_\al^{\rm max})\over 2mg\sin\al}
=1+\O(B) \quad{\rm as}\ B\to0 
\eeq
We have used {\sl Surface evolver} to compute the ratio 
(\ref{ratioB}) numerically for $\al=30^\circ$ and $\al=60^\circ$ 
as function of volume, when the contact angle at $g=0$
is $\theta_0\simeq 32^\circ$, see Fig. \ref{gg} (right), where
$B=17.2\times(V/100)^{(2/3)}$.

Equations (\ref{ode000})(\ref{ode01}) can be solved exactly.
In equation (\ref{ode01}) the change of variable $t=r^2/R^2$ and function
$v=rz_{11}/R^2$ leads to 
\beqn
(1-t)^{-1/2}=-4(1-t)v''+6v'-2{v\over t}\,,\quad v(0)=0,\ v(t_0)=0
\eeqn
{\sl Mathematica} gives the solution:
\beqa\label{v}
v=t(1-t)^{-1/2}\,C_1+{1\over3}-{1\over3}(1-t)^{1/2}\hskip1cm\cr
+{1\over3}t(1-t)^{-1/2}\ln\bigl(1+(1-t)^{1/2}\bigr)
\eeqa
where $C_1$ is fixed by $v(t_0)=0$.
\begin{figure}
\begin{center}
\vskip-3cm
\resizebox{8cm}{!}{\includegraphics{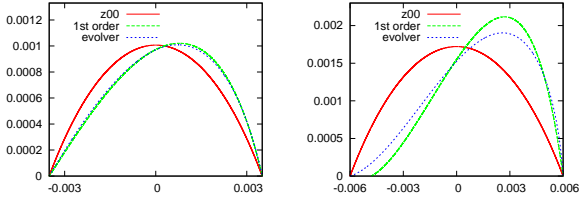}} 
\caption{Drop on $30^\circ$ incline: spherical cap $z_{00}$, first order 
$z_{00}+Bz_1$, and {\sl Surface evolver} profiles. Right: volume 100$\mu$L,
base radius $r_0=6\,$mm, Bond number $B=17.2$. Left: volume 20$\mu$L, base 
radius $6\times 0.2^{1/3}\,$mm, Bond number $B=5.89$. Abscissa $x$ along incline,
downwards, and ordinate $z$ perpendicular to incline, both in meters.}
\label{drop}
\end{center}
\end{figure}
Equation (\ref{ode000}) was solved in \cite{F}. We give here an equivalent 
solution: 
\beqn
z_{01}''+r^{-1}(1-t)^{-1}(1-4t)z_{01}'=R^{-2}(1-t)^{-3/2}(z_{00}-Rp_1)
\eeqn
or, with $u=z_{01}'r/R$ and $t_0=r_0^2/R^2$,
\beqn
2\,u'-{3u\over1-t}=(1-t)^{-1}-\bigl((1-t_0)^{1/2}+p_1\bigr)(1-t)^{-3/2}
\eeqn
This is a first order linear differential equation which can be solved by the
variation of constants method, yielding
\beqn
u(t)={1\over3}(1-t)^{-3/2}-{1\over3}-\12\bigl((1-t_0)^{1/2}+p_1\bigr)\,t(1-t)^{-3/2}
\eeqn
The volume of the drop doesn't vary:
\beqn
0=2\pi\int_0^{r_0}dr\,rz_{01}=-\pi\int_0^{r_0}dr\,r^2z_{01}'
=-{\pi R^3\over2}\int_0^{t_0}dt\,u
\eeqn
implying
\beq
p_1={{8\over3}-2t_0+(1-t_0)^{1/2}\Bigl({2t_0\over3}-{8\over3}\Bigr)\over
2-t_0-2(1-t_0)^{1/2}}
\eeq
Then
\beq\label{z1}
z_{01}(r)=\int_{r_0}^rd\ell\, z_{01}'(\ell)={R\over2}\int_{t_0}^tds\,{u(s)\over s}
={R\over2}\Bigl({I\over3}+J\Bigr)
\eeq
where
\beqn
I=2(1-t)^{-1/2}-2(1-t_0)^{-1/2}
-2\log\biggl[{1+(1-t)^{1/2}\over1+(1-t_0)^{1/2}}\biggr]
\eeqn
\beqn
J=\bigl((1-t_0)^{1/2}+p_1\bigr)\,\bigl((1-t_0)^{-1/2}-(1-t)^{-1/2}\bigr)
\eeqn

Resulting $z=z_{00}+Bz_{01}\cos\al+Bz_{11}\sin\al\,\cos\phi$ with $z_{01}$ given by
(\ref{z1}) and  $z_{11}=v R^2/r$ given by (\ref{v}) are shown on Fig. \ref{drop} 
as ``first order'', together with the spherical cap $z_{00}$ and the almost exact
{\sl Surface evolver} results. The profile with $B=17.2$, corresponding to the 
top right points for the ratio on Fig. \ref{gg} (left and right figures), is not
far from the physical limitation $\theta^{\rm min}_\al=0$,
and the first order approximation in fact gives a small negative value for
$\theta^{\rm min}_\al$. 

\subsection{Overhangs}
\begin{figure}
\begin{center}
\vskip-2.6cm
\resizebox{8.5cm}{!}{\includegraphics{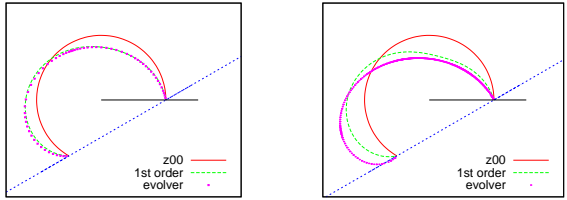}} 
\caption{Drop on $30^\circ$ hydrophobic incline: spherical cap $z_{00}$
with contact angle $2\pi/3$, first order $z_{00}+Bz_1$ and {\sl Surface
evolver} profiles. Left: Bond number B = 0.5, corresponding
to a volume $V\simeq25\mu$L and base radius $r_0\simeq 1.7$mm. Right:
B = 0.8, corresponding to $V\simeq51\mu$L and $r_0\simeq 2.1$mm. The
profiles are scaled by a factor $r_0^{-1}$ for comparison.}
\label{drop2pi3}
\end{center}
\end{figure}

The derivation so far used height functions $z(r,\phi)$, which excludes 
overhangs and contact angles larger than $\pi/2$. Yet singularities only
occur at contact angles 0 and $\pi$, beyond which a fraction of the drop profile
would go into $z<0$ if continued analytically. The laws (\ref{mga})(\ref{mgaB}) 
therefore extend to 
$0<\theta^{\rm min}<\theta^{\rm max}<\pi$. As for the drop profile,
the apparent singularity at $\pi/2$ disappears in polar coordinates with origin
at the center of the spherical cap for $B=0$, angle $\theta\in[0,\theta_0]$ 
measured from the $z$-axis and angle $\phi\in[0,2\pi]$ as before,
\beqn
\vec r\,(\theta,\phi)=\bigl(R+\del r(\theta,\phi)\bigr)\vec e_r\,,\qquad 
\del r(\theta_0,\phi)=0 
\eeqn
New formulas are derived from the old, first in the case $\theta_0<\pi/2$ using
$$\del r=B z_{01}\cos\al\cos\theta+Bz_{11}\sin\al\cos\theta\cos\phi$$ and 
$(1-t_0)^{1/2}=\cos\theta_0$, $(1-t)^{1/2}=\cos\theta$ and then extended
analytically to the whole range of $\theta$ with $\theta_0\in]0,\pi[$,
where the cosines can be negative. Results are shown on Fig. \ref{drop2pi3}.
One may note that the first order in $B$
overestimates the effect of gravity in the hydrophilic case
but underestimates in the hydrophobic case.

The scope of the present study is the physical range
$0\le\theta^R\le\theta^{\rm min}_\al\le\theta^{\rm max}_\al\le\theta^A\le\pi$.
It is remarkable that the first order approximation is 
typically within 1\% of the almost exact {\sl Surface evolver} result in this 
full range, whatever the advancing and receding contact angles 
$\theta^A$ and $\theta^R$. 
The exact solution of the linearized 
Laplace-Young equation given in the present work together with the simple 
formulas (\ref{mga})(\ref{mgaB}) should therefore be valuable.

\begin{acknowledgments}
This research was partially funded by the Inter-University Attraction Poles 
Programme (IAP 7/38 MicroMAST) of the Belgian Science Policy Office. The authors
also thank FNRS and R\'egion Wallonne for partial support.
\end{acknowledgments}


\parskip 1pt
\baselineskip 1pt
\begin{thebibliography}{00}

\bibitem{BT} V. Berejnov, R. E. Thorne: {\sl
Effect of transient pinning on stability of drops sitting on an inclined plane.}
Phys. Rev. E {\bf 75}, 066308 (2007).

\bibitem{Be} J. Berthier: {\sl Microdrops and digital microfluidics.}
Elsevier, second edition (2013).

\bibitem{BB} J. Berthier, K. Brakke: {\sl The physics of microdroplets.}
Scrivener-Wiley (2012).

\bibitem{Br} K. Brakke: {\sl Surface Evolver Manual.} Susquehanna University
(2013). 

\bibitem{BOS} R. A. Brown, F. M. Orr, Jr., L. E. Scriven: {\sl
Static Drop on an Inclined Plate: Analysis by the Finite Element Method.}
J. Colloid Interface Sci. {\bf 73}, 76-87 (1980).

\bibitem{CHST} T.-H. Chou, S.-J. Hong, Y.-J. Sheng, H.-K. Tsao:
{\sl Drops sitting on a tilted plate: Receding and advancing pinning.}
Langmuir {\bf 28}, 5158-5166 (2012).

\bibitem{EJ04I}  A.I. ElSherbini, A.M. Jacobi: 
{\sl Liquid drops on vertical and inclined 
surfaces I. An experimental study of drop geometry.}
 J. Colloid Interface Sci.  {\bf 273}, 556--565 (2004).

\bibitem{EJ06} A.I. ElSherbini, A.M. Jacobi: 
{\sl Retention forces and contact angles for critical liquid
          drops on non-horizontal surfaces.}
J. Colloid Interface Sci. {\bf 299}, 841 (2006).

\bibitem{F} A.H. Fatollahi: {\sl 
On the shape of a lightweight drop on a horizontal plane.} 
Phys. Scr. {\bf 85} (2012) 045401 (6pp)

\bibitem{Fr} Y. I. Frenkel: {\sl On the behavior of liquid drops on a solid 
surface 1. The sliding of drops on an inclined surface.}
 Zh. Eksp. Teor. Fiz. {\bf 18}, 659 (1948). 
          Translated by V. Berejnov: http://xxx.lanl.gov/abs/physics/0503051

\bibitem{dGBQ} P.-G. de Gennes, F. Brochard-Wyart, D. Qu\'er\'e: {\sl
Capillarity and Wetting Phenomena: Drops, Bubbles,
Pearls, Waves.} Springer-Verlag (2003).

\bibitem{MO} G. Macdougall, C. Ockrent: {\sl Surface energy relations in 
liquid/solid systems.} Proc. R. Soc. London, Ser. A {\bf 180}, 151-173 (1942).

\bibitem{MSKK} M. Musterd, V. van Steijn, C. R. Kleijn, and M. T. Kreutzer: {\sl
Droplets on inclined plates: Local and global hysteresis of pinned capillary 
surfaces.} Phys. Rev. Lett. {\bf 113}, 066104 (2014).

\bibitem{WSRV} J. A. White, M. J. Santos, M. A. Rodr\'iguez-Valverde, 
S. Velasco: {\sl Numerical Study of the Most Stable Contact Angle of Drops on 
Tilted Surfaces.} Langmuir {\bf 31}, 5326-5332 (2015).
\end{thebibliography}
\end{document}